%% file: IP_StochCLS-ljamt2.tex
\documentclass[copyright,creativecommons]{eptcs}
\providecommand{\event}{}
\providecommand{\volume}{}
\providecommand{\anno}{2009}
\providecommand{\firstpage}{1}

\usepackage{breakurl}
\usepackage{amsmath, xcolor}
\usepackage{amssymb}
\usepackage{amsfonts}
\usepackage{epsfig}
\usepackage{prooftree}

\usepackage{float}
\usepackage{afterpage}
\restylefloat{figure}

\input{macros}

\title{Modelling an Ammonium Transporter with SCLS\thanks{This research is founded by the BioBITs Project (\emph{Converging
Technologies} 2007, area: Biotechnology-–ICT), Regione Piemonte.}}
\author{Mario Coppo$^1$, Ferruccio Damiani$^1$, Elena Grassi$^{1,2}$, Mike Guether$^3$ and Angelo Troina$^1$
\institute{$^1$Dipartimento di Informatica, Universit\`a di Torino}
\institute{$^2$Molecular Biotechnology Center, Dipartimento di Genetica, Biologia e Biochimica, Universit\`a di Torino}
\institute{$^3$Dipartimento di Biologia Vegetale, Universit\`a di Torino}
\email{\{coppo,damiani,troina\}@di.unito.it $\quad$ grassi.e@gmail.com $\quad$ mike.guether@unito.it}
}
\def\titlerunning{Modelling an Ammonium Transporter with SCLS}
\def\authorrunning{M. Coppo, F. Damiani, E. Grassi, M. Guether and A. Troina}
\begin{document}
\maketitle

\begin{abstract}
The Stochastic Calculus of Looping Sequences (SCLS) is a recently
proposed modelling language for the representation and simulation of
biological systems behaviour. It has been designed with the aim of
combining the simplicity of notation of rewrite systems with the
advantage of compositionality. It also allows a
rather simple and accurate description of biological membranes and
their interactions with the environment.

In this work we apply SCLS to model a newly discovered ammonium
transporter. This transporter is believed to play a fundamental
role for plant mineral acquisition, which takes place in the
arbuscular mycorrhiza, the most wide-spread plant-fungus symbiosis
on earth. Due to its potential application in agriculture this
kind of symbiosis is one of the main focuses of the BioBITs
project.

%[OLD: This transporter is differentially expressed in arbuscular cells and is believed to play a fundamental role in the nutrients uptake which takes place in the context of plants-fungi symbiosis (namely, the \emph{Lotus japonicus} plant upon colonization with the \emph{Gigaspora margarita} fungus) at the plant roots level.]

In our experiments the passage of $NH_3$ / $NH_4^+$ from the
fungus to the plant has been dissected in known and hypothetical
mechanisms; with the model so far we have been able to simulate
the behaviour of the system under different conditions. Our
simulations confirmed some of the latest experimental results
about the LjAMT2;2 transporter. The initial simulation results of
the modelling of the symbiosis process are promising and indicate
new directions for biological investigations.

\end{abstract}

\input{SecIntro}

\input{SecCLS}

\input{IP_SecModel}

\input{SecDiscussion}

\input{SecConc}

\paragraph{Acknowledgements} We thank the referees for their insightful
comments. The final version of the paper improved due to their
suggestions.

\bibliographystyle{plain}
\bibliography{fmb,pure_bio}

\end{document}

%% file: macros.tex
\newcommand{\qqop}[1]{\mathrel{\makebox[2em]{$#1$}}}

\newcommand{\agr}{\quad\big|\quad}

\newcommand{\mycdot}{\!\cdot\!}
\newcommand{\CC}{\mathcal{C}}

\newcommand{\EE}{\mathcal{E}}

\newcommand{\RR}{\mathcal{R}}
\newcommand{\PP}{\mathcal{P}}
\newcommand{\Seq}{\mathcal{S}}
\newcommand{\XX}{\mathcal{X}}
\newcommand{\TT}{\mathcal{T}}

\newcommand{\OO}{\mathcal{O}}
\newcommand{\SSq}{\mathcal{S}}

\newcommand{\VV}{\mathcal{V}}
\newcommand{\xx}{\widetilde{x}}
\newcommand{\yy}{\widetilde{y}}
\newcommand{\zz}{\widetilde{z}}

\newcommand{\ltrans}[1]{\xrightarrow{#1}}

\newcommand{\Loop}[1]{\left(#1\right)^L}

\newcommand{\rewrites}{\; \mapsto \;}
\newcommand{\srewrites}[1]{\stackrel{#1}{\mapsto}}

\newcommand{\into}{\ensuremath{\,\rfloor}\,}
\newcommand{\pipe}{\ensuremath{\,|\,}}
\newcommand{\phole}{\square}

\newcommand{\bbbr}{{\rm I\!R}}

%% file: SecIntro.tex
\section{Introduction}

Given the central role of agriculture in worldwide economy, several
ways to optimize the use of costly artificial fertilizers are now
being actively pursued. One approach is to find methods to nurture
plants in more ``natural'' manners, avoiding the complex chemical
production processes used today. In the last decade the Arbuscular
Mycorrhiza (AM), the most widespread symbiosis between plants and
fungi, got into the focus of research because of its potential as a
natural plant fertilizer. Briefly, fungi help plants to acquire
nutrients as phosphorus (P) and nitrogen (N) from the soil whereas
the plant supplies the fungus with energy in form of
carbohydrates~\cite{Par08}. The exchange of these nutrients is
supposed to occur mainly at the eponymous arbuscules, a specialized
fungal structure formed inside the cells of the plant root. The
arbuscules are characterized by a juxtaposition of a fungal and a
plant cell membrane where a very active interchange of nutrients is
facilitated by several membrane transporters. These transporters are
surface proteins that facilitate membrane crossing of molecules
which, because of their inherent chemical nature, are not freely
diffusible.

As long as almost each cell in the majority of multicellular
organisms shares the same genome, modern theories point out that
morphological and functional differences between them are mainly
driven by different genes expression~\cite{MBOC1}. Thanks to the
last experimental novelties~\cite{TB99,NCB01} a precise analysis of
which genes are expressed in a single tissue is possible; therefore
is possible to identify genes that are pivotal in specific
compartments and then study their biological function. Following
this route a new membrane transporter has been discovered by
expression analysis and further characterized~\cite{GUE09}. This
transporter is situated on the plant cell membrane which is directly
opposed to the fungal membrane, located in the arbuscules. Various
experimental evidence points out that this transporter binds to an
$NH_{4}^+$ moiety outside the plant cell, deprotonates it, and
mediates inner transfer of $NH_{3}$, which is then used as a
nitrogen source, leaving an $H^+$ ion outside. The AM symbiosis is
far from being unraveled: the majority of fungal transporters and
many of the chemical gradients and energetic drives of the symbiotic
interchanges are unknown. Therefore, a valuable task would be to
model \emph{in silico} these conditions and run simulations against
the experimental evidence available so far about this transporter.
Conceivably, this approach will provide biologists with working
hypotheses and conceptual frameworks for future biological
validation.

In computer science, several formalisms have been proposed for the description of the behaviour of biological systems.
Automata-based models~\cite{ABI01,MDNM00} have the advantage of allowing the direct use of many verification tools such as model checkers.
Rewrite systems~\cite{DL04,P02,BMMT06,ABSTT07} usually allow describing biological systems with a notation that can be easily understood by biologists.
Compositionality allows studying the behaviour of a system componentwise.
Both automata-like models and rewrite systems present, in general, problems from the point of view of compositionality, which,
instead, is in general ensured by process calculi, included those commonly used to describe biological systems~\cite{RS02,PRSS01,Car05}.

The Stochastic Calculus of Looping Sequences (SCLS)~\cite{BMMTT08} (see also~\cite{BMMT06,BMMT06s,M07}) is a recently proposed modelling language for the representation and simulation of biological systems behaviour. It has been designed with the aim of combining the simplicity of notation of rewrite systems with the advantage of a form of compositionality. It also allows a rather simple and accurate description of biological membranes and their interactions with the environment.

In this work we apply SCLS to model the mentioned transporter. This transporter is differentially expressed in arbuscular cells and is believed to play a fundamental role in the nutrients uptake which takes place in the context of plants-fungi symbiosis at the root level. This symbiosis is one of the main focuses of the BioBITs project, due to its relevant role in agriculture.

On these premises, the aim of this work is to
model the interchange between the fungus-plant interface
and the plant cells, using as a reference system the $NH_{3}$/$NH_{4}^+$ turnover.
This could disclose which is the driving
power of the net nitrogen flux inside plant cells (still unknown at the
chemical level). Furthermore, this information may also be exploited to model other,
and so far poorly characterized, transporters.

\paragraph{Outline}
Section~\ref{CLS_formalism} introduces the syntax and the semantics of the SCLS and shows some modelling guidelines. Section~\ref{SecModel} presents the SCLS representations of the ammonium transporter and our experimental results, discussed in Section~\ref{SecDiscussion}. Section~\ref{SecConc} concludes by outlining some further work.

%% file: SecCLS.tex
\section{The Stochastic Calculus of Looping Sequences}\label{CLS_formalism}

In this section we briefly recall the Stochastic Calculus of Looping
Sequences (SCLS)~\cite{BMMTT08}. A SCLS (biological) model consists
of a term, representing the structure of the modelled system and a
set of stochastic rewrite rules which determine its evolution.

\paragraph{Terms}%\label{CLS_syntax}

Terms of SCLS are defined from a set of atomic symbols
representing the basic elements of the system under description,
which could correspond, according to the level of the description,
to genes, proteins, molecules or even single chemical elements.
Terms are built from the atomic elements via the operators of
sequencing (defining an ordered sequence of atomic elements),
looping (defining membranes from looping sequences) and parallel
composition (representing coexistence in the same biological
ambient).

Assuming a possibly infinite alphabet $\EE$ of symbols ranged over
by $a,b,c,\ldots$  the syntax of  \emph{SCLS terms} $T$ and
\emph{sequences} $S$ is defined by the grammar:
\begin{align*}
T\; & \qqop{::=} \; S \!\agr\! \Loop{S} \into T \!\agr\! T \pipe T\\
S\; & \qqop{::=} \; \epsilon \!\agr\! a  \!\agr\! S \mycdot S
\end{align*}
\normalsize where $a$ is any element of $\EE$ and $\epsilon$ is the
empty sequence. We denote the infinite sets of terms and sequences
with $\TT$ and $\Seq$, respectively.
%\end{definition}

The SCLS terms are built from the elements of $\EE$ by means of four
syntactic operators: sequencing $\_\cdot\_$,  looping $\Loop{\_}$,
parallel composition $\_\pipe\_$, and the containment operator
$\_\into\_$.  Sequencing is used to concatenate atomic elements. The
looping operator transforms a sequence in a closed loop when
combined with
 another
term via the $\_\into\_$ operator.
 A term can then be either a sequence, or a looping sequence
containing another term, or the parallel composition of two terms.
By the definition of terms, we have that looping and containment are
always associated, hence we can consider $\Loop{\_} \into \_$ as a
single binary operator that applies to one sequence and one term.

The biological interpretation of the operators is the following: the
main entities which occur in cells are DNA and RNA strands,
proteins, membranes, and other macro-molecules. DNA strands (and
similarly RNA strands) are sequences of nucleic acids, but they can
be seen also at a higher level of abstraction as sequences of genes.
Proteins are sequences of amino acids which typically have a very
complex three-dimensional structure. In a protein there are usually
(relatively) few subsequences, called domains, which actually are
able to interact with other entities by means of chemical reactions.
SCLS sequences can model DNA/RNA strands and proteins by describing
each gene or each domain with a symbol of the alphabet. Membranes
are closed surfaces often interspersed with proteins, and may have a
content. A closed surface can be modelled by a looping sequence. The
elements (or the subsequences) of the looping sequence may represent
the proteins on the membrane, and by the containment operator it is
possible to specify what the membrane contains. Other
macro-molecules can be modeled as single alphabet symbols, or as
sequences of their components. Finally, juxtaposition of entities
can be described by the parallel composition operator of their
representations. More detailed description of the biological
interpretation of the SCLS operators involved in the description of
the system presented in this paper will be given
%in Section~\ref{sec:modelguide}.
at the end of this section.
%together with some modelling
%guidelines
%A deeper description of the biological
%interpretation of CLS operators together with some modelling
%guidelines will be given in Section~\ref{sec:modelguide}.

\begin{figure}[t]
\begin{center}
\begin{minipage}{0.98\textwidth}
\begin{center}
\includegraphics[height=40mm]{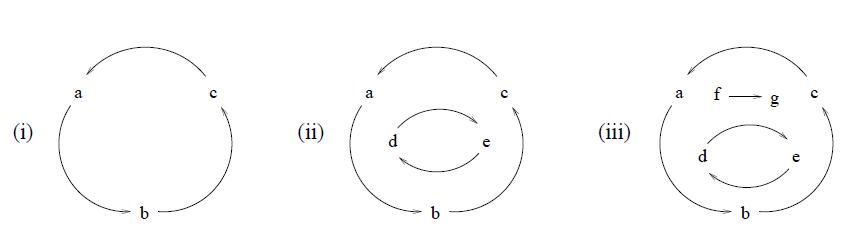} %%%%**** was 30mm
\end{center}
\vspace{-0.5cm}
\caption{(i) $\Loop{a
\mycdot b \mycdot c} \into \epsilon$; (ii) $\Loop{a
\mycdot b \mycdot c} \into \Loop{d \mycdot e} \into \epsilon$; (iii) $\Loop{a
\mycdot b
\mycdot c} \into ((\Loop{d \mycdot e} \into \epsilon) \pipe f \mycdot g)$.}
\label{fig:loop_seq_fig_CLS}
\end{minipage}
\end{center}
\end{figure}

Brackets can be used to indicate the order of application of the
operators. We assume that $\Loop{\_} \into \_$ has the
precedence over $\_\pipe\_$ and $\_\cdot\_$ over $\_\pipe\_$, therefore $  \Loop{S} \into
T_1 \pipe T$ has to be read as $(\Loop{S} \into T_1) \pipe T $ and
$a \cdot b \pipe c$ as $(a\cdot b)\pipe c$. An example of an SCLS term is $ a
\pipe b \pipe \Loop{m \cdot n} \into (c \cdot d \pipe e) $
consisting of three entities $a$, $b$ and $\Loop{m \cdot n} \into (c
\cdot d \pipe e)$. It represents a membrane with two molecules $m$
and $n$ (for instance, two proteins) on its surface, and containing
a sequence $c \cdot d$ and a molecule
$e$. Molecules $a$ and $b$ are outside the membrane. See
Figure~\ref{fig:loop_seq_fig_CLS} for some graphical
representations.

%\subsection{Semantics}\label{CLS_semantics}

\paragraph{Structural congruence} As usual in systems of this kind, SCLS
terms representing the same entity may have syntactically different structures. Such terms are thus identified by
means of a \emph{structural congruence} relation. In particular
the structural congruence relations $\equiv_S$ and $\equiv_T$ are
defined as the least congruence relations on sequences and on
terms, respectively, satisfying the following rules:
% \small
\begin{gather*}
S_1 \mycdot ( S_2 \mycdot S_3 )
    \equiv_S ( S_1 \mycdot S_2 ) \mycdot S_3 \qquad
S \mycdot \epsilon \equiv_S \epsilon \mycdot S
    \equiv_S S
\\[-0.1\baselineskip]
S_1 \equiv_S S_2 \mbox{ implies } S_1 \equiv_T S_2 \mbox{ and }
    \Loop{S_1} \into T \equiv_T \Loop{S_2} \into T\\[-0.1\baselineskip]
T_1 \pipe T_2 \equiv_T T_2 \pipe T_1 \quad T_1 \pipe ( T_2 \pipe T_3
) \equiv_T (T_1 \pipe T_2) \pipe T_3 \quad
T \pipe \epsilon \equiv_T T \\[-0.1\baselineskip]
T_1 \equiv_T T_2 \mbox{ implies } %
    \Loop{S} \into T_1 \equiv_T \Loop{S} \into T_2\\[-0.1\baselineskip]
\Loop{\epsilon} \into \epsilon \equiv_T \epsilon \qquad \Loop{S_1
\mycdot S_2} \into T \equiv_T \Loop{S_2 \mycdot S_1} \into T
\end{gather*}
%\normalsize

Structural congruence states the associativity of $\cdot$ and
$\pipe$, the commutativity of the latter and the neutral role of
$\epsilon$. Moreover, axiom $\Loop{S_1 \mycdot S_2} \into T \equiv_T
\Loop{S_2 \mycdot S_1} \into T$ says that looping sequences can
rotate. In the following we will simply use $\equiv$ in place of
$\equiv_T$, $\equiv_S$.

\paragraph{Rewrite Rules, Variables and Patterns}
A rewrite rule is defined as a pair of terms (possibly containing
variables), which represent the patterns defining the system
transformations, together with a rate representing the speed of
the modelled reaction. Rules are applicable to all subterms,
identified by the notion of reduction context introduced below,
which match the left-hand side of the rule via a proper
instantiation of its variables. The system transformation is
obtained by replacing the reduced subterm by the corresponding
instance of the right-hand side of the rule.

Variables in patterns can be of three kinds: two are associated
with the two different syntactic categories of terms and
sequences, and one is associated with single alphabet elements. We
assume a set of term variables $TV$ ranged over by $X,Y,Z,\ldots$,
a set of sequence variables $SV$ ranged over by
$\xx,\yy,\zz,\ldots$, and a set of element variables $\XX$ ranged
over by $x,y,z,\ldots$. All these sets are pairwise disjoint and
possibly infinite. We denote by $\VV$ the set of all variables $TV
\cup SV \cup \XX$, and with $\rho$ any variable in $\VV$. A
pattern is a term which may include variables, i.e.
\emph{patterns} $P$ and \emph{sequence patterns} $SP$ of SCLS are
given by the following grammar:
% \small
\begin{align*}
P\; & \qqop{::=} \; SP \agr \Loop{SP} \into P \agr P \pipe P \agr X\\
SP\; & \qqop{::=} \;  \epsilon \agr a  \agr SP \mycdot SP \agr \xx \agr x
\end{align*}
\normalsize where $a$ is an element of $\EE$, and $X,\xx$ and $x$
are elements of $TV,SV$ and $\XX$, respectively. The infinite set
of patterns is denoted with $\PP$.
 The structural congruence
relation can be be trivially extended to patterns. \\ An
\emph{instantiation} is a partial function $\sigma : \VV \rightarrow
\TT$ which preserves the type of variables, thus for $X \in TV,
\xx \in SV$ and $x \in \XX$ we must have $\sigma(X) \in \TT,
\sigma(\xx) \in \Seq$, and $\sigma(x) \in \EE$, respectively. Given
$P \in \PP$, the expression $P \sigma$ denotes the term obtained by
replacing each occurrence of each variable $\rho \in \VV$ appearing
in $P$ with the corresponding term $\sigma(\rho)$.

Let $\Sigma$ denote the set of all the possible instantiations and
$Var(P)$  the set of variables appearing in $P \in \PP$.
%
%\begin{definition}[Stochastic Rewrite Rule]
%A {\em stochastic rewrite rule} is a triple $(P_1,P_2,k)$, denoted
%with $P_1 \! \srewrites{k} \! P_2$, where $P_1,P_2 \in \PP$, $P_1
%\not\equiv \epsilon$ and such that $Var(P_2) \subseteq Var(P_1)$; $k
%\in \bbbr^{\geq 0}$ is the {\em rewrite rate}.
%\end{definition}
%
Then a \emph{rewrite rule} is a triple $(P_1,P_2, k)$, denoted
with $P_1 \! \srewrites{k} \! P_2$, where $k \in \bbbr^{\geq 0}$
is the kinetic constant of the modeled chemical reaction, $P_1,P_2
\in \PP$, $P_1 \not\equiv \epsilon$ such that $Var(P_2) \subseteq
Var(P_1)$. A rewrite rule $P_1 \! \srewrites{k} \! P_2$ then
states that a subterm $P_1 \sigma$, obtained by instantiating
variables in $P_1$ by some instantiation function $\sigma$, can be
transformed into the subterm $P_2\sigma$ with kinetics given by
the constant $k$.

\paragraph{Contexts}
The definition of reduction for SCLS systems is completed by
resorting to the notion of reduction context. To this aim, as
usual, the syntax of terms is enriched with a new element $\phole$
representing a hole. \emph{Reduction context} (ranged over by $C$)
are defined by:
$$
C \;\; ::= \; \; \phole \agr C \pipe T \agr T \pipe C \agr \Loop{S} \into C
$$
where $T \in \TT$ and $S \in \SSq$. The context $\phole$ is called
the \emph{empty context}. We denote with $\CC$ the infinite set of
contexts.

By definition, every context contains a single $\phole$. Let us
assume $C,C'\in \CC$. Then $C[T]$ denotes the term obtained by
replacing $\phole$ with $T$ in $C$; similarly $C[C']$ denotes
context composition, whose result is the context obtained by
replacing $\phole$ with $C'$ in $C$. The structural equivalence is extended to
contexts in the natural way (i.e. by considering $\phole$ as a new
and unique symbol of the alphabet $\EE$).

Note that the general form of rewrite rules does not permit to have
sequences as contexts. A rewrite rule introducing a parallel
composition on the right hand side (as $a \srewrites{k} b \pipe c$)
applied to an element of a sequence (e.g., $m \mycdot a \mycdot m$)
would result into a syntactically incorrect term (in this case $m
\cdot (b\pipe c) \cdot m$); which moreover does not seem to have
 any biological meaning, parallel and sequence represent orthogonal
 notions.
Therefore to modify a sequence, a pattern representing the whole
sequence must appear in the rule. For example, rule $a \mycdot \xx
\srewrites{k} a \pipe \xx$ can be applied to any sequence starting
with element $a$, and, hence, the term $a\mycdot b$ can be rewritten
as $a \pipe b$, and the term $a \mycdot b \mycdot c$ can be
rewritten as $a \pipe b \mycdot c$. The notion of reduction
context will forbid that a substitution like this is applied to
looping sequences.

\paragraph{Stochastic Reduction Semantics}
The operational semantics of SCLS is defined by incorporating a
collision-based stochastic framework along the line of the one
presented by Gillespie in \cite{G77}, which is, \emph{de facto}, the
standard way to model quantitative aspects of biological systems.
Following the law of mass action, it is necessary to count the
number of reactants that are present in a system in order to compute
the exact rate of a reaction. The same approach has been
applied, for instance, to the stochastic $\pi$-calculus
\cite{P95,PRSS01}. The idea of Gillespie's algorithm is that a rate
constant is associated with each considered chemical reaction. Such
a constant is obtained by multiplying the kinetic constant of the
reaction by the number of possible combinations of reactants that
may occur in the system. The resulting rate is then used as the
parameter of an exponential distribution modelling the time spent
between two occurrences of the considered chemical reaction.

The use of exponential distributions to represent the (stochastic)
time spent between two occurrences of chemical reactions allows
describing the system as a Continuous Time Markov Chain (CTMC), and
consequently allows verifying properties of the described system
analytically and by means of stochastic model checkers.

The number of reactants in a reaction represented by a reduction
rule is evaluated considering the number of occurrences of subterms
to which the rule can be applied and of the terms produced. For
instance in evaluating the application rate of the rewrite rule $R=
a \pipe b \srewrites{k} c$ to the term $T=a\pipe a \pipe b \pipe b$
we must consider the number of the possible combinations of
reactants of the form $a\pipe b$ in $T$. Since each occurrence of
$a$ can react with each occurrence of $b$, this number is 4. So the
application rate of $R$ is $k\cdot 4$.

The evaluation of the application rate of a reduction rule
containing variables is more complicate since there can be many
different ways in which variables can be instantiated to match the
subterm to be reduced, and this must be considered to correctly
evaluate the application rate. The technique to do this is
described in~\cite{BMMTT08}. With this technique, given two terms
$T,T'$ and a reduction rule $R$, we can compute the function
$\OO(R,T,T')$, defining the number of the possible applications of
the rule $R$ to the term $T$ resulting in the term $T'$. We refer
to~\cite{BMMTT08} for more details and explanation.

Given a finite set $\RR$ of stochastic rewrite rules, then, the
\emph{reduction semantics} of SCLS is the least labelled transition
relation satisfying the following rule:
%\small
$$
\frac{ R=P_1 \srewrites{k}  P_2 \in \RR
        \quad T \equiv C[P_1\sigma]  \quad T'\equiv C[P_2 \sigma]}
    { T \ltrans{k\cdot \OO(R,T,T')}  T'}
$$
\normalsize
%\end{definition}
 The rate of the reduction is then obtained as
the product of the rewrite rate constant and the number of
occurrences of the rule within the starting term (thus counting the
exact number of reactants to which the rule can be applied and which
produce the same result).  The rate associated with each transition
in the stochastic reduction semantics is the parameter of an
exponential distribution that characterizes the stochastic behaviour
of the activity corresponding to the applied rewrite rule. The
stochastic semantics is essentially a \emph{Continuous Time Markov
Chain} (CTMC). A standard simulation procedure that corresponds to
Gillespie's simulation algorithm~\cite{G77} can be followed. The
most recent implementation of SCLS, based on Gillespie's algorithm,
is described in \cite{Sca07}.

\paragraph{Modelling Guidelines}%\label{sec:modelguide}

%\begin{table}[tbp]
%\begin{center}
%%\small
%\begin{tabular}{|l|l|}
%\hline
%{\bf Biomolecular Entity} & {\bf SCLS Term} \\
%\hline
%\hline
%Elementary object & Alphabet symbol\\
%(genes, domains, & \\
% other molecules, etc...) & \\
%\hline
%DNA strand
%    & Sequence of elements repr. genes \\
%\hline
%RNA strand
%    & Sequence of elements repr.
%     transcribed genes \\
%\hline
%Protein
%    & Sequence of elements repr. domains\\
%    & or single habet symbol\\
%\hline
%Molecular population
%    & Parallel composition of molecules \\
%\hline
%Membrane
%    & Looping sequence\\
%\hline
%\end{tabular}
%\end{center}
%\caption{Guidelines for the abstraction of biomolecular entities
%into CLS.}\label{tab:guidelines-entities}
%\end{table}

\begin{table}[tbp]
\begin{center}
%\small
\begin{tabular}{|l|l|}
\hline
{\bf Biomolecular Event} & {\bf Examples of SCLS Rewrite Rule} \\
\hline \hline State change &
    $a \rewrites b$ \\
    & $\xx \cdot a \cdot \yy \rewrites \xx \cdot b \cdot \yy$ \\
\hline Complexation &
    $a \pipe b \rewrites c$ \\
    & $\xx \cdot a \cdot \yy  \pipe b \rewrites \xx \cdot c \cdot \yy$ \\
\hline Decomplexation &
    $c \rewrites a \pipe b$ \\
    & $\xx \cdot c \cdot \yy  \rewrites \xx \cdot a \cdot \yy \pipe b$ \\
\hline Catalysis &
    $c \pipe P_1 \rewrites c \pipe P_2$ %\\
    %&
    \qquad (where $P_1 \rewrites P_2$ is the catalyzed event) \\
\hline
%State change &
%    $\Loop{a \cdot \xx} \into X \rewrites \Loop{b \cdot \xx} \into X$ \\
%on membrane
%    &  \\
%\hline
%Complexation &
%    $\Loop{a \cdot \xx \cdot b \cdot \yy} \into X \rewrites
%    \Loop{c \cdot \xx \cdot \yy} \into X $ \\
%on membrane
%    &  $ a \pipe \Loop{ b \cdot \xx} \into X \rewrites
%    \Loop{c \cdot \xx} \into X $ \\
%    &  $ \Loop{ b \cdot \xx} \into (a \pipe  X) \rewrites
%    \Loop{c \cdot \xx} \into X $ \\
%\hline
%Decomplexation &
%    $ \Loop{c \cdot \xx } \into X \rewrites
%    \Loop{a \cdot b \cdot \xx} \into X$ \\
%on membrane
%    &  $ \Loop{c \cdot \xx} \into X \rewrites
%    a \pipe \Loop{ b \cdot \xx} \into X$ \\
%    &  $ \Loop{c \cdot \xx} \into X \rewrites
%    \Loop{ b \cdot \xx} \into (a \pipe  X) $ \\
%\hline
%Catalysis &
%    $\Loop{c \cdot \xx \cdot SP_1 \cdot \yy} \into X\rewrites
%    \Loop{c \cdot \xx \cdot SP_2 \cdot \yy} \into X$ \\
%on membrane
%    & where $SP_1 \rewrites SP_2$ is the catalyzed event \\
%\hline
Membrane crossing &
    $a \pipe \Loop{\xx} \into X \rewrites
    \Loop{\xx} \into (a \pipe X)$ \\
    & $\Loop{\xx} \into (a \pipe X) \rewrites
    a \pipe \Loop{\xx} \into X$ \\
    & $\xx \cdot a \cdot \yy \pipe \Loop{\zz} \into X \rewrites
    \Loop{\zz} \into (\xx \cdot a \cdot \yy \pipe X)$ \\
    & $\Loop{\zz} \into (\xx \cdot a \cdot \yy \pipe X) \rewrites
    \xx \cdot a \cdot \yy \pipe \Loop{\zz} \into X$ \\
\hline Catalyzed&
    $a \pipe \Loop{b \cdot \xx} \into X \rewrites
    \Loop{b \cdot \xx} \into (a \pipe X)$ \\
membrane crossing &
    $\Loop{b \cdot \xx} \into (a \pipe X) \rewrites
    a \pipe \Loop{b \cdot \xx} \into X$ \\
    & $\xx \cdot a \cdot \yy \pipe \Loop{b \cdot \zz} \into X \rewrites
    \Loop{b \cdot \zz} \into (\xx \cdot a \cdot \yy \pipe X)$ \\
    & $\Loop{b \cdot \zz} \into (\xx \cdot a \cdot \yy \pipe X) \rewrites
    \xx \cdot a \cdot \yy \pipe \Loop{b \cdot \zz} \into X$ \\
\hline Membrane joining &
    $\Loop{\xx} \into (a \pipe X) \rewrites \Loop{a \cdot \xx} \into X$ \\
    & $\Loop{\xx} \into (\yy \cdot a \cdot \zz \pipe X) \rewrites
    \Loop{\yy \cdot a \cdot \zz \cdot \xx} \into X$\\
\hline Catalyzed &
    $\Loop{b \cdot \xx} \into (a \pipe X)
    \rewrites \Loop{a \cdot b \cdot \xx} \into X$ \\
%membrane joining &
%    $\Loop{\xx} \into (a \pipe b \pipe X) \rewrites \Loop{a \cdot \xx}
%    \into (b \pipe X)$ \\
%    & $\Loop{b \cdot \xx} \into (\yy \cdot a \cdot \zz \pipe X)
%    \rewrites \Loop{\yy \cdot a \cdot \zz \cdot \xx} \into X$\\
%    & $\Loop{\xx} \into (\yy \cdot a \cdot \zz \pipe b \pipe X)
%    \rewrites \Loop{\yy \cdot a \cdot \zz \cdot \xx} \into (b \pipe X)$\\
%\hline
%Membrane fusion &
%    $\Loop{\xx} \into (X) \pipe \Loop{\yy} \into (Y) \rewrites
%    \Loop{\xx \cdot \yy} \into (X \pipe Y)$ \\
%\hline
%Catalyzed &
%    $\Loop{a \cdot \xx} \into (X) \pipe \Loop{b \cdot \yy} \into
%(Y)\rewrites $\\
%membrane fusion &
%    $ \qquad \Loop{a \cdot \xx \cdot b\cdot \yy} \into (X \pipe Y)$
%\\
%\hline
%Membrane division &
%    $\Loop{\xx \cdot \yy} \into (X \pipe Y) \rewrites
%    \Loop{\xx} \into (X) \pipe \Loop{\yy} \into (Y)$ \\
%\hline
%Catalyzed &
%    $\Loop{a \cdot \xx \cdot b\cdot \yy} \into (X \pipe Y) \rewrites$\\
%membrane division &
%    $\qquad \Loop{a \cdot \xx} \into (X) \pipe \Loop{b \cdot \yy} \into (Y)$
%\\
\hline
\end{tabular}
\normalsize
\end{center}
\caption{Guidelines for the abstraction of biomolecular events
into SCLS.}\label{tab:guidelines-events}
\end{table}

SCLS can be used to model biomolecular systems analogously to what
is done, e.g,  by Regev and Shapiro in \cite{RS04} for the
$\pi$-calculus. An abstraction is a mapping from a real-world
domain to a mathematical domain, which may allow highlighting some
essential properties of a system while ignoring other,
complicating, ones. In \cite{RS04}, Regev and Shapiro show how to
abstract biomolecular systems as concurrent computations by
identifying the biomolecular entities and events of interest and
by associating them with concepts of concurrent computations such
as concurrent processes and communications.
%In particular, they give some guidelines for the
%abstraction of biomolecular systems to the $\pi$-calculus, and give
%some simple examples.

The use of rewrite systems, such as SCLS, to describe biological
systems is founded on a different abstraction. Usually, entities
(and their structures) are abstracted by terms of the rewrite
system, and events by rewrite rules. We have already recalled the
biological interpretation of SCLS operators in the previous
section.

In order to describe cells, it is quite natural to consider
molecular populations and membranes. Molecular populations are
groups of molecules that are in the same compartment of the cell.
As we have said before, molecules can be of many types: they could
be classified as DNA and RNA strands, proteins, and other
molecules.
%DNA and RNA strands
%and proteins can be seen as non-elementary objects. DNA strands are
%composed by genes, RNA strands are composed by parts corresponding
%to the transcription of individual genes, and proteins are composed
%by parts having the role of interaction sites (or domains). Other
%molecules are considered as elementary objects, even if they are
%complexes.
Membranes are considered as elementary objects, in the sense that
we do not describe them at the level of the lipids they are made
of. The only interesting properties of a membrane are that it may
have a content (hence, create a compartment) and that it may have
molecules on its surface.

We give now some examples of biomolecular events of interest and
their description in SCLS. The simplest kind of event is the change
of state of an elementary object. Then, we consider interactions
between molecules: in particular complexation, decomplexation and
catalysis. These interactions may involve single elements of
non-elementary molecules (DNA and RNA strands, and proteins).
Moreover, there can be interactions between membranes and molecules:
in particular a molecule may cross or join a membrane.

Table~\ref{tab:guidelines-events} lists the guidelines (taken
from~\cite{BMMTT08}) for the abstraction into SCLS rules of the
biomolecular events we will use in our
application.\footnote{Kinetics are omitted from the table for
simplicity.} Entities are associated with SCLS terms: elementary
objects are modelled as alphabet symbols, non-elementary objects
as SCLS sequences and membranes as looping sequences. Biomolecular
events are associated with SCLS rewrite rules.

%% file: IP_SecModel.tex
\section{Modelling the Ammonium Transporter with SCLS}\label{SecModel}

The scheme in Figure~\ref{fig:symbschem} (taken from~\cite{GUE09})
illustrates nitrogen, phosphorus and carbohydrate exchanges at the
mycorrhizal interface according to previous works and the results
of~\cite{GUE09}. In this paper we focus our investigation on the
sectors labelled with \textbf{(c)}, \textbf{(a1)} and \textbf{(a2)}.
Namely, we will present SCLS models for the equilibrium between
$NH_4^+$ and $NH_3$ and the uptake by the LjAMT2;2 transporter
\textbf{(c)}, and the exchange of $NH_4^+$ from the fungus to the
interspatial level \textbf{(a1-2)}. The choice of SCLS is
motivated by the fact that membranes, membrane elements (like
LjAMT2;2) and the involved reactions can be represented in it in a
quite natural way.

The simulations illustrated in this section are done with the
SCLSm prototype simulator~\cite{Sca07}. In the following we will
use a more compact notation for the parallel composition operator,
namely, we will write $a\times n$ to denote the parallel
composition of $n$ atomic elements $a$. Moreover, to simplify the
counting mechanism, we might use different names for the same
molecule when it belongs to different compartments. For instance,
occurrences of the molecule $NH_3$ inside the plant cell are
called $NH_3\textrm{\_inside}$.

\afterpage{\clearpage\begin{figure}[H]
\begin{center}
\begin{minipage}{0.98\textwidth}
\begin{center}
\includegraphics[height=140mm]{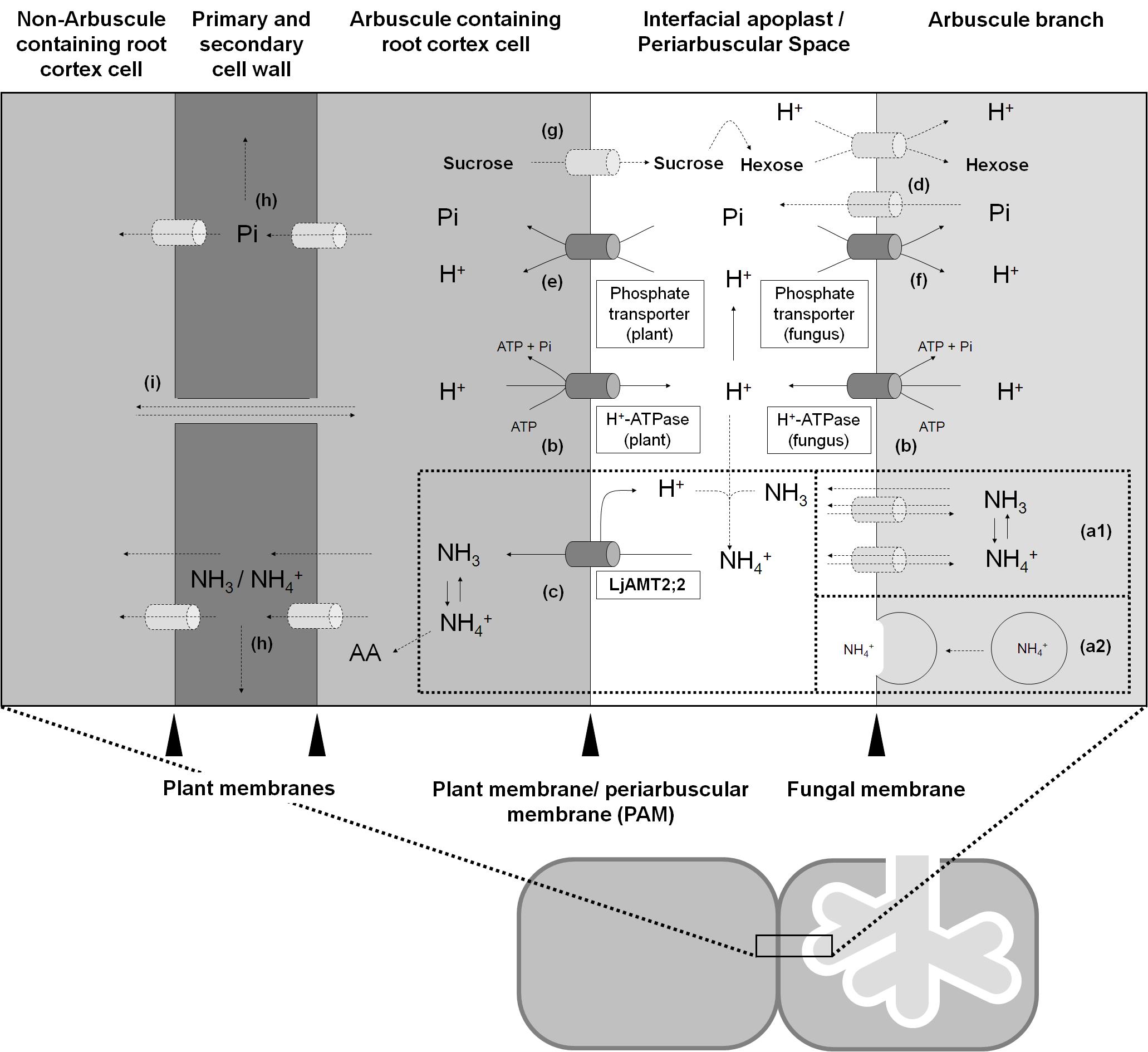}
\end{center}
\vspace{0.5cm}
\caption{\footnotesize \textbf{(a1-2)} $NH_3/NH_4^+$ is released in the arbuscules from arginine which is transported from the extra- to the intraradical fungal structures~\cite{Gov05}. $NH_3$/$NH_4^+$ is released by so far unknown mechanisms (transporter, diffusion \textbf{(a1)} or vesicle-mediated \textbf{(a2)}) into the periarbuscular space (PAS) where, due to the acidic environment, its ratio shifts towards $NH_4^+$ ($>99.99\%$). \textbf{(b)} The acidity of the interfacial apoplast is established by plant and fungal $H^+$-ATPases~\cite{Hau05, Bal07} thus providing the energy for $H^+$-dependent transport processes. \textbf{(c)} The $NH_4^+$ ion is deprotonated prior to its transport across the plant membrane via the LjAMT2;2 protein and released in its uncharged $NH_3$ form into the plant cytoplasm. The $NH_3$/$NH_4^+$ acquired by the plant is either transported into adjacent cells or immediately incorporated into amino acids (AA). \textbf{(d)} Phosphate is released by so far unknown transporters into the interfacial apoplast. \textbf{(e)} The uptake of phosphate on the plant side then is mediated by mycorrhiza-specific Pi-transporters~\cite{JPH07, GUE09}. \textbf{(f)} AM fungi might control the net Pi-release by their own Pi-transporters which may reacquire phosphate from the periarbuscular space~\cite{Bal07}. \textbf{(g)} Plant derived carbon is released into the PAS probably as sucrose and then cleaved into hexoses by sucrose synthases~\cite{Hon03} or invertases~\cite{SRH06}. AM fungi then acquire hexoses~\cite{Sha95, Sol97} and transport them over their membrane by so far unknown hexose transporters. It is likely that these transporters are proton co-transporter as the GpMST1 described for the glomeromycotan fungus Geosiphon pyriformis~\cite{NT00}. Exchange of nutrients between arbusculated cells and non-colonized cortical cells can occur by apoplastic \textbf{(h)} or symplastic \textbf{(i)} ways.}
\label{fig:symbschem}
\end{minipage}
\end{center}
\end{figure}}%ENDafterpage

\subsection{$NH_3$/$NH_4^+$ Equilibrium}
\begin{figure}[t]
\begin{center}
\begin{minipage}{0.98\textwidth}
\begin{center}
\includegraphics[width=80mm]{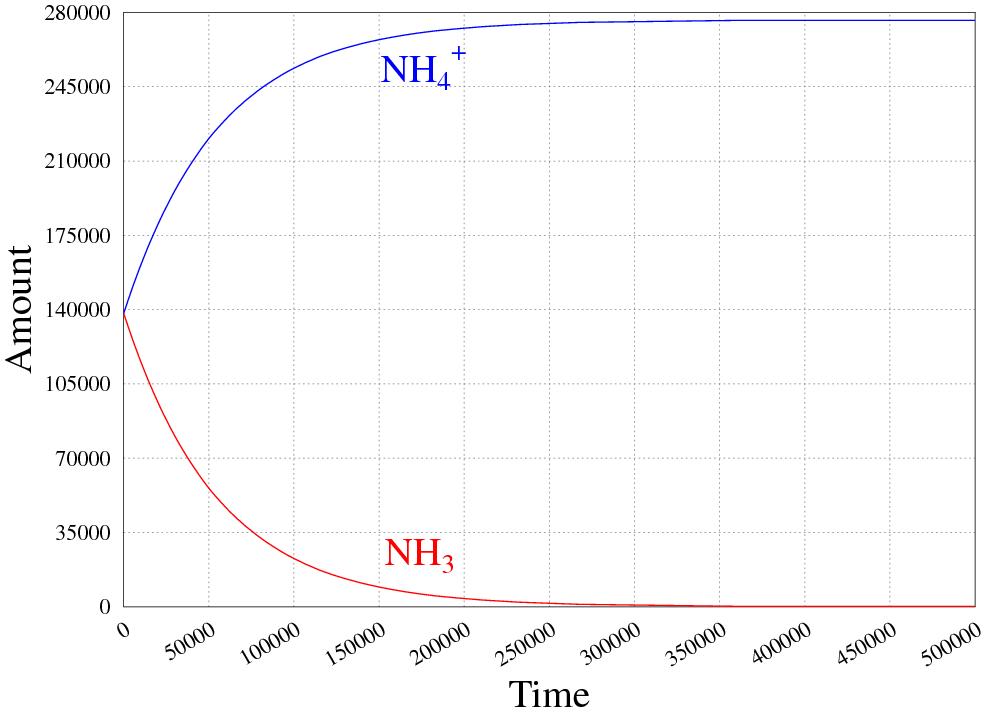}
\end{center}
\vspace{-0.5cm}
\caption{Extracellular equilibrium between $NH_3$ and $NH_4^+$.}
\label{fig:NH3eqNH4}
\end{minipage}
\end{center}
\end{figure}

We decided to
start modelling a simplified pH equilibrium, at the interspatial level (right part of section \textbf{(c)} in Figure~\ref{fig:symbschem}),
without
considering $H_2O$, $H^+$ and $OH^-$; therefore we tuned the reaction rates in
order to reach the correct percentages of $NH_3$
over total $NH_3/NH_4^+$ in the different compartments.
Like these all the rates and initial terms used in this work are obtained by manual adjustments made
looking at the simulations results and trying to keep simulations times acceptable - we plan to refine these rates
and numbers in future works to reflect more deeply the available biological data.
Following~\cite{GUE09}, we consider an extracellular pH of 4.5~\cite{Gut00}.
In such conditions, the percentage of molecules of $NH_3$ over the sum $NH_3 + NH_4^+$ should be around $0.002$.
The reaction we considered is the following:
$$
NH_3 \overset{k_1}{\underset{k_2}{\rightleftharpoons}}
NH_4^+
$$
with $k_1=0.018*10^{-3}$ and $k_2=0.562*10^{-9}$. One can
translate this reaction with the SCLS rules.
\begin{equation*}\tag*{(R1)}
NH_3 \srewrites{k_1} NH_4^+
\end{equation*}
\begin{equation*}\tag*{(R2)}
NH_4^+ \srewrites{k_2} NH_3
\end{equation*}
In Figure~\ref{fig:NH3eqNH4} we show the results of this first simulation given the initial term $T=NH_3\times 138238 \pipe NH_4^+\times 138238$.

This equilibrium is different at the intracellular level (pH around 7 and 8)~\cite{GBD92}, so we use two new rules to model the transformations of $NH_3$ and $NH_4^+$ inside the cell, namely:
\begin{equation*}\tag*{(R3)}
NH_3\textrm{\_inside} \srewrites{k_1} NH_4^+\textrm{\_inside}
\end{equation*}
\begin{equation*}\tag*{(R4)}
NH_4^+\textrm{\_inside} \srewrites{k_2'} NH_3\textrm{\_inside}
\end{equation*}
where $k_2'=0.562*10^{-6}$.

\subsection{LjAMT2;2 Uptake}

We can now present the SCLS model of the uptake of the LjAMT2;2
transporter (left part of section \textbf{(c)} in
Figure~\ref{fig:symbschem}). We add a looping sequence modelling
an arbusculated plant cell. Since we are only interested in the
work done by the LjAMT2;2 transporter, we consider a membrane
containing this single element. The work of the transporter is
modelled by the rule:
\begin{equation*}\tag*{(R5)}
NH_4^+ \pipe \Loop{LjAMT2} \into (X) \srewrites{k_t} H^+ \pipe
\Loop{LjAMT2}\into (X \pipe NH_3\textrm{\_inside})
\end{equation*}
where $k_t=0.1*10^{-5}$.

%So, starting from the term $T=NH3\times 138238 \pipe NH_4^+\times 138238 \pipe \Loop{LjAMT2}\into \epsilon$, we obtain %the results in Figure~\ref{fig:nh4_ljamt2}, where we show the concentrations of $NH_3$ and $NH_4^+$ inside the cell, and %the number of $NH_4^+$ molecules brought in by the transporter (BrIn).
%
%\begin{figure}[t]
%\begin{center}
%\begin{minipage}{0.98\textwidth}
%\begin{center}
%\includegraphics[height=58mm]{Simulations/nh4_ljamt2.jpg}
%\includegraphics[height=58mm]{Simulations/zoom_nh4_ljamt2.jpg}
%\end{center}
%\caption{At equal concentrations.}
%\label{fig:nh4_ljamt2}
%\end{minipage}
%\end{center}
%\end{figure}

\begin{figure}[t]
\begin{center}
\begin{minipage}{0.98\textwidth}
\begin{center}
\includegraphics[width=77mm]{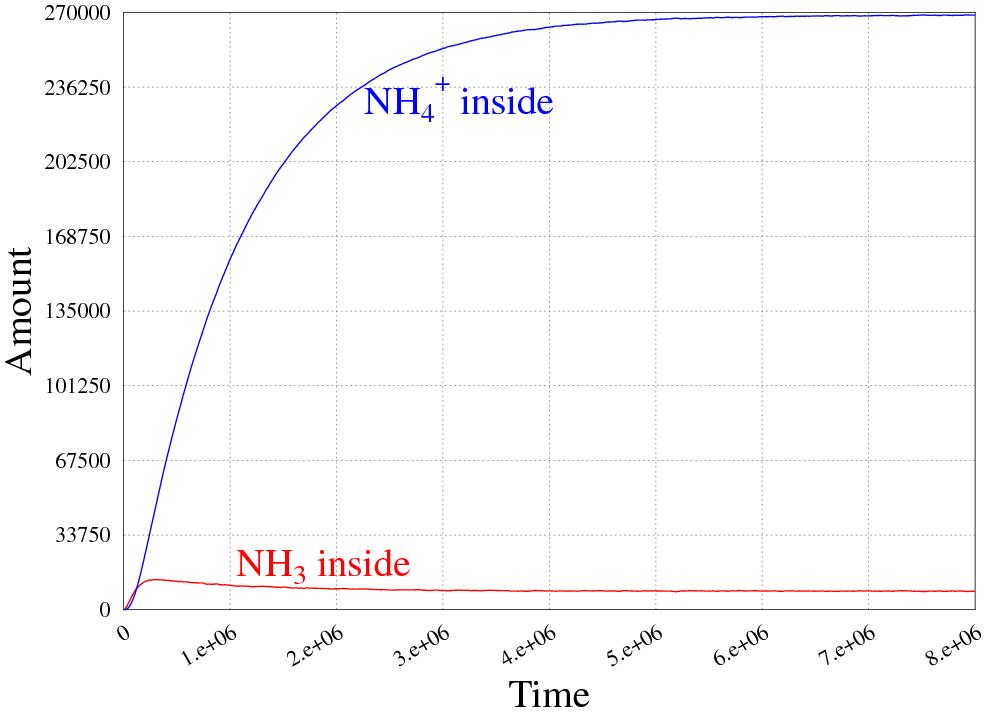}
\includegraphics[width=77mm]{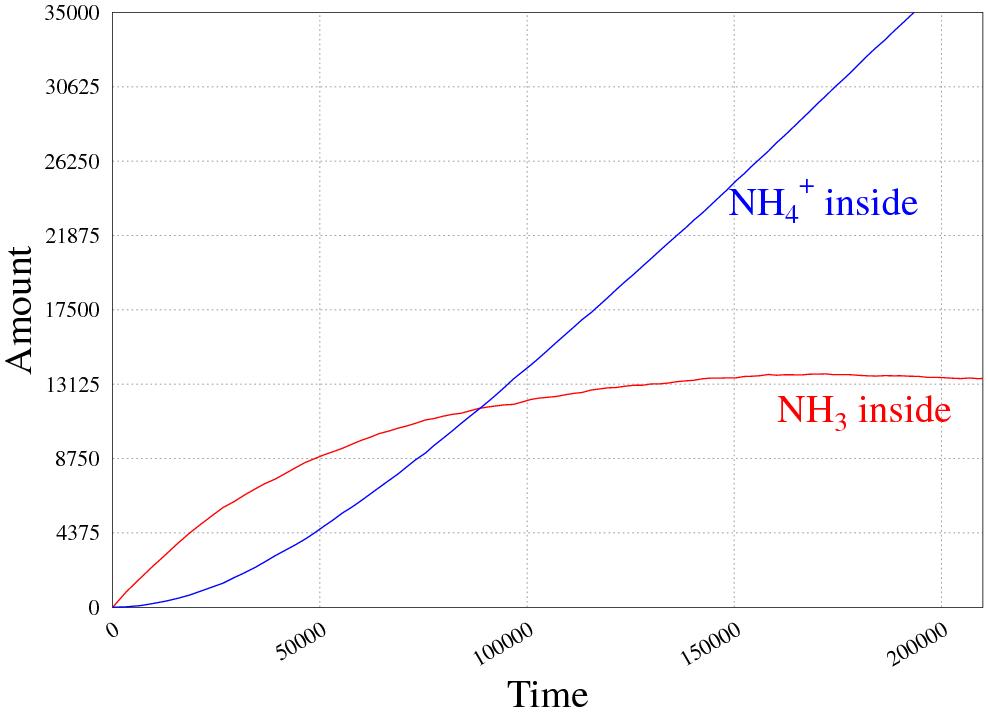}
\end{center}
\caption{At high $NH_4^+$ concentration.}
\label{fig:nh4_ljamt2_noNH3}
\end{minipage}
\end{center}
\end{figure}

\begin{figure}[t]
\begin{center}
\begin{minipage}{0.98\textwidth}
\begin{center}
\includegraphics[width=77mm]{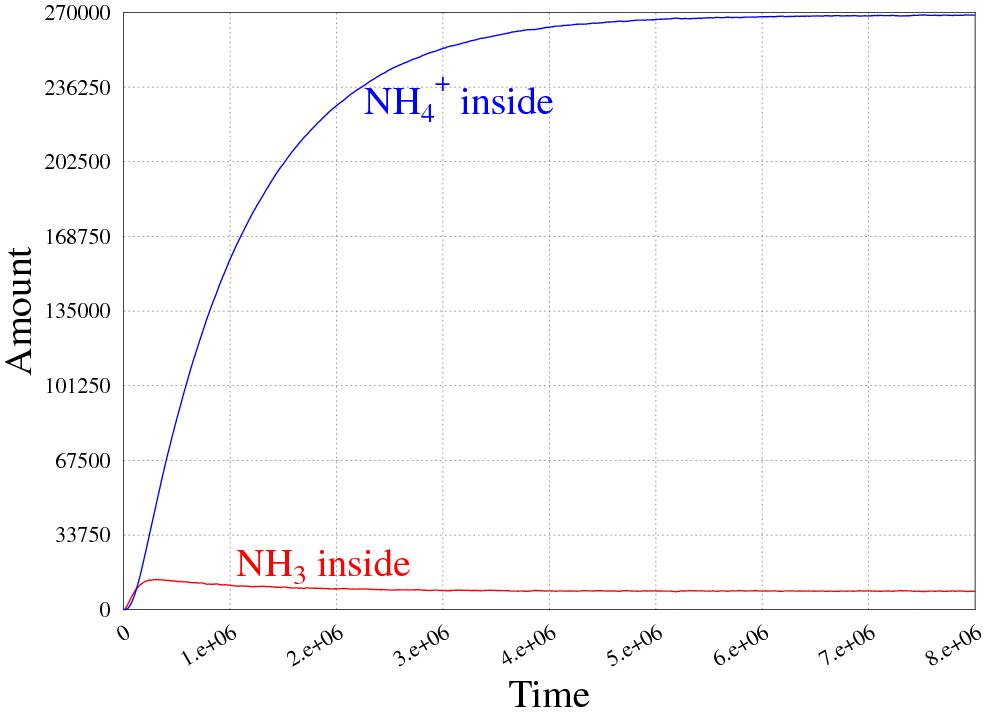}
\includegraphics[width=77mm]{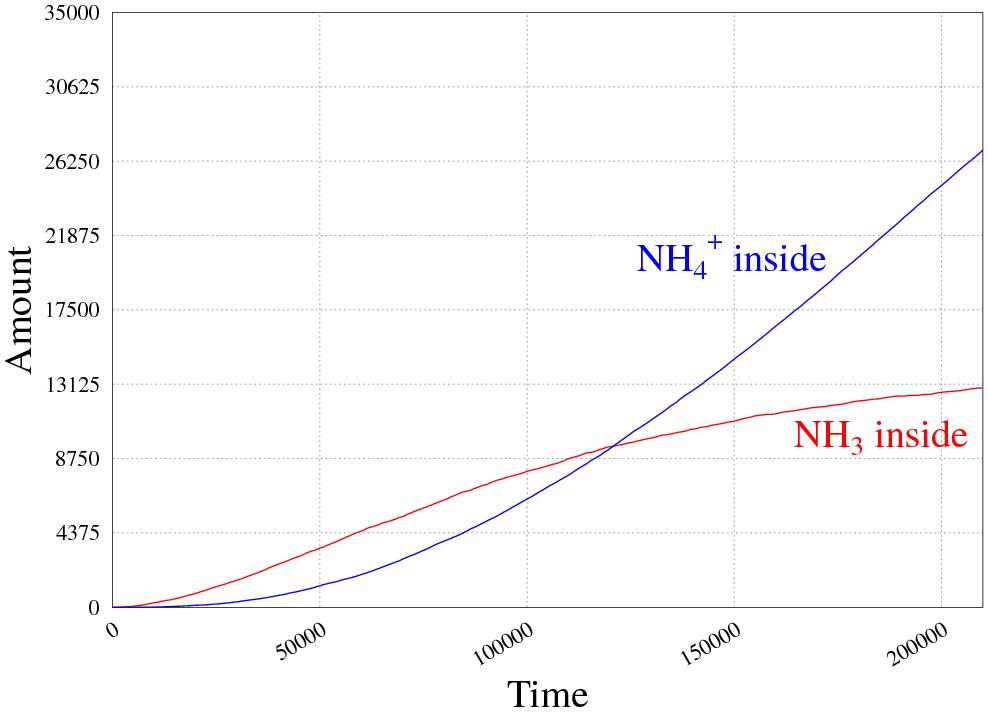}
\end{center}
\vspace{-0.5cm}
\caption{At low $NH_4^+$ concentration.}
\label{fig:nh4_ljamt2_noNH4}
\end{minipage}
\end{center}
\end{figure}

We can investigate the uptake rate of the transporter at different
initial concentrations of $NH_3$ and $NH_4^+$.
Figure~\ref{fig:nh4_ljamt2_noNH3} and
Figure~\ref{fig:nh4_ljamt2_noNH4} show the results for the initial
terms\\$T_1=NH_3\times 776 \pipe NH_4^+\times 276400 \pipe
\Loop{LjAMT2}\into \epsilon$ and $T_2=NH_3\times276400 \pipe
NH_4^+\times776 \pipe \Loop{LjAMT2}\into \epsilon$, respectively
(the left one represents the whole simulations, while on the right
there is a magnification of their initial segment).


\begin{figure}[t]
\begin{center}
\begin{minipage}{0.98\textwidth}
\begin{center}
\includegraphics[width=77mm]{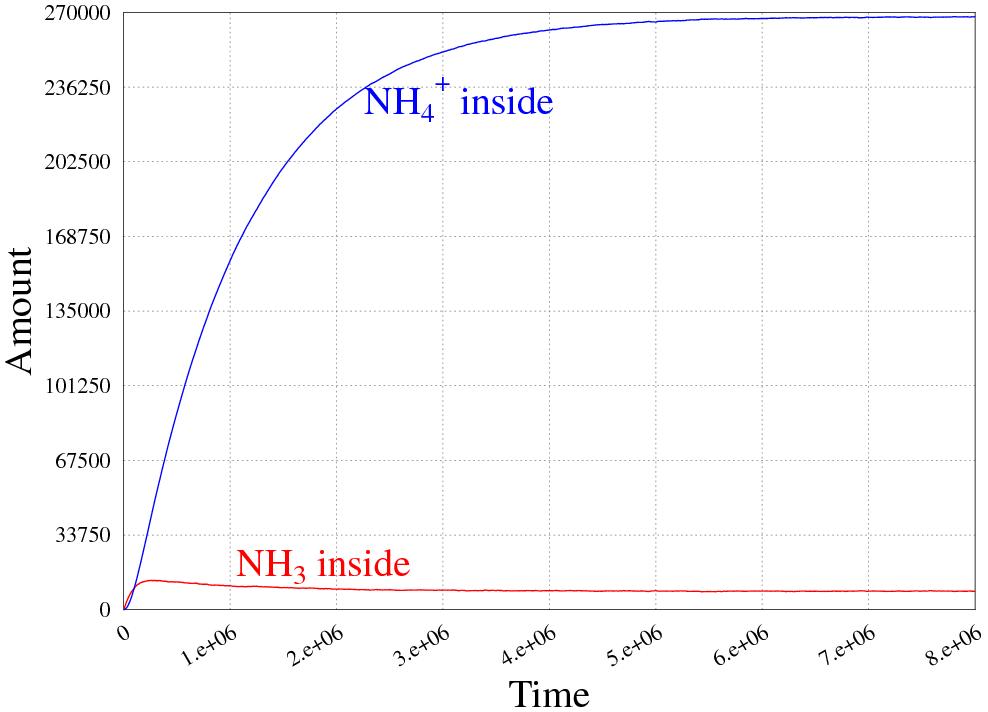}
\includegraphics[width=77mm]{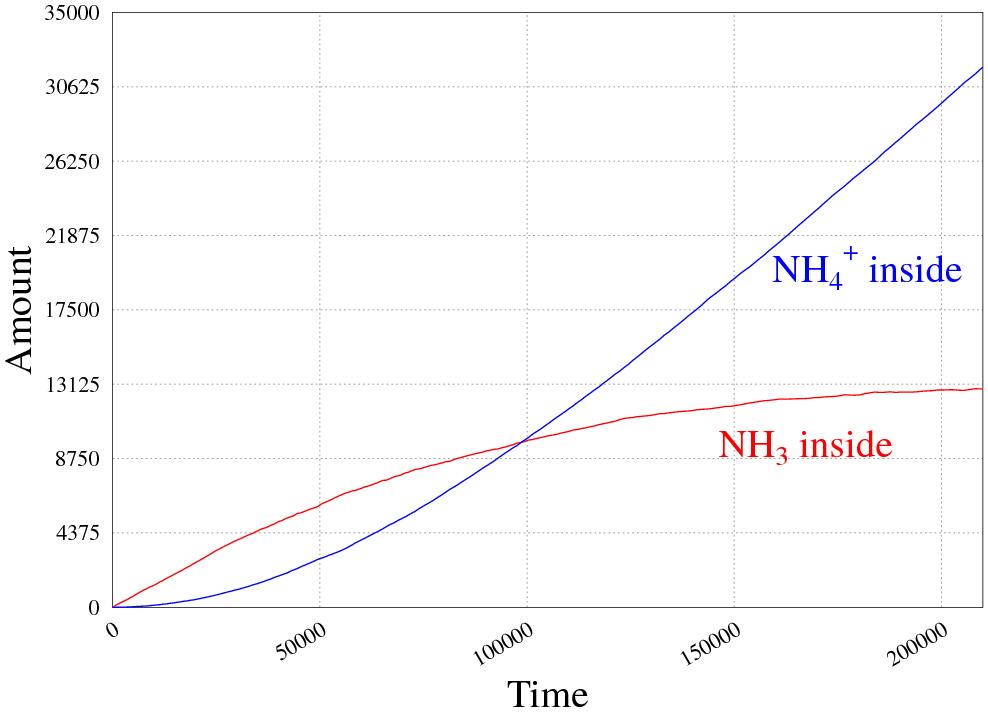}
\end{center}
\vspace{-0.5cm}
\caption{At extracellular pH=7.}
\label{fig:nh4_ljamt2_pH7}
\end{minipage}
\end{center}
\end{figure}

We can also investigate the uptake rate of the transporter at
different extracellular pH. Namely, we consider an extracellular pH
equal to the intracellular one (pH around 7 and 8), obtained by
imposing $R1$ and $R2$ equal to $R3$ and $R4$, respectively, i.e. $k_2 = k_2'$.
Figure~\ref{fig:nh4_ljamt2_pH7} shows the results for the initial
terms $T_1=NH_3\times 138238 \pipe NH_4^+\times 138238 \pipe
\Loop{LjAMT2}\into \epsilon$.

\begin{figure}[t]
\begin{center}
\begin{minipage}{0.98\textwidth}
\begin{center}
\includegraphics[width=77mm]{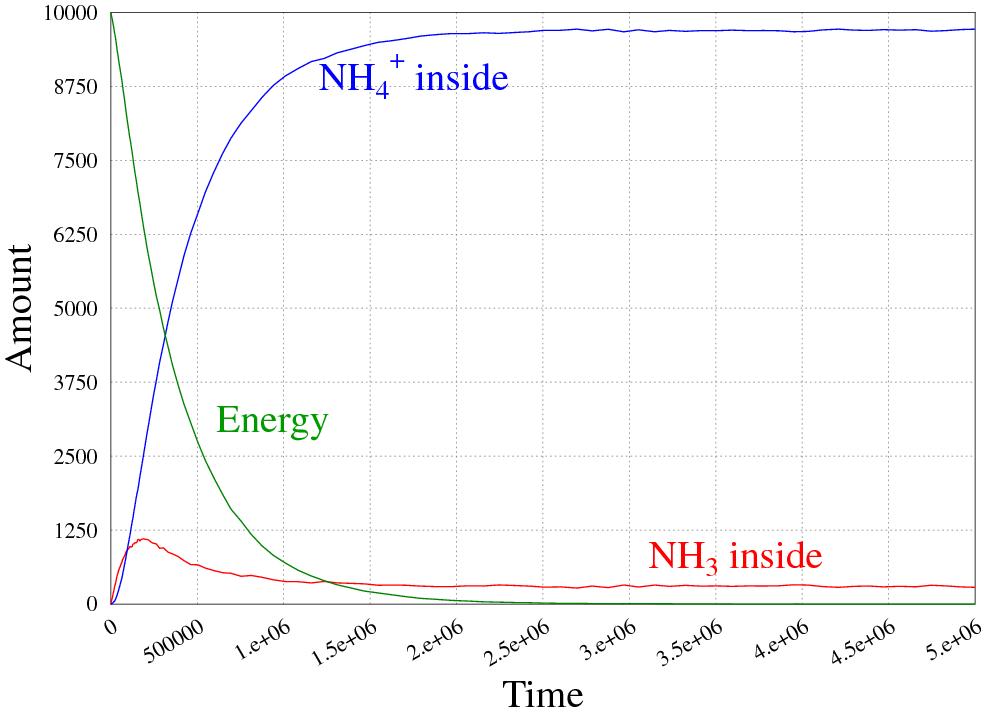}
\includegraphics[width=77mm]{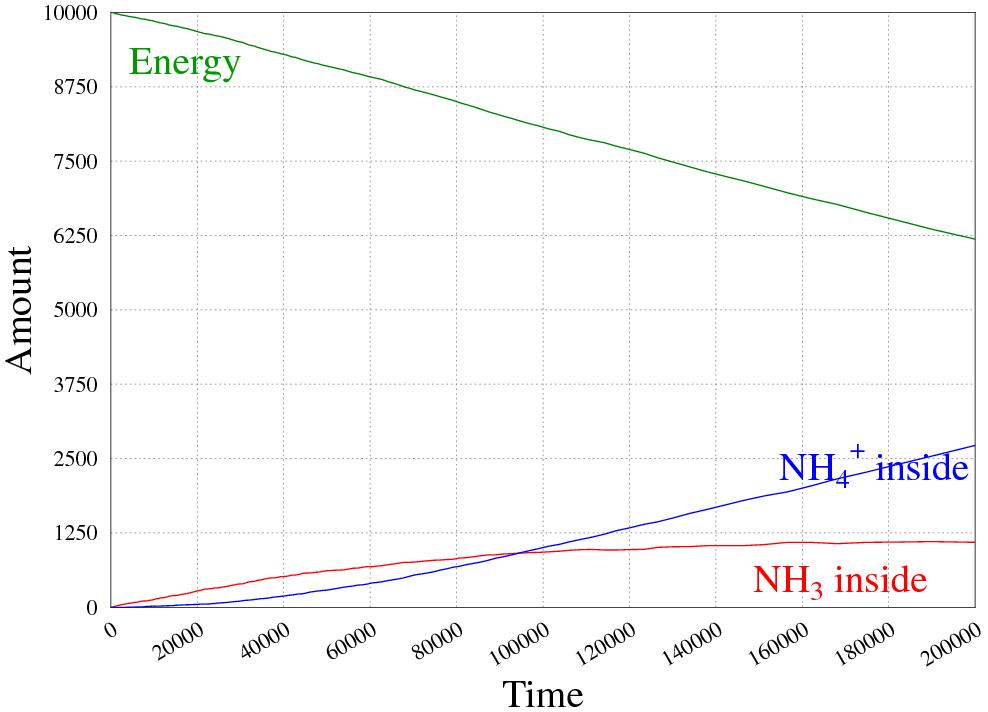}
\end{center}
\vspace{-0.5cm}
\caption{LjAMT2;2 with active energy.}
\label{fig:nh4_ljamt2_energy}
\end{minipage}
\end{center}
\end{figure}

%This models the work of the transporter if no energy is required to do the actual work. What happens,
%instead, if the transport actually requires some energy, as it happens in many real cases. We might suppose that the
%cell actually contains a certain quantity of this energy. Since we are only considering a possible hypothesis we do not need %to specify here in which form this energy is going to be provided (we just suppose that the transporter works only in the
%presence of some other particle which takes a role similar to the one of ATP in many pumps). Therefore, we can modify rule
%R5 modelling the transporter role with the following one:
Since now we modeled the transporter supposing that no active form of energy is required to do the actual work -
which means that the $NH_4^+$ gradient between the cell and the extracellular ambient is sufficient to
determine a net uptake.
The predicted tridimensional structure of LjAMT2;2 suggests that it
does not use ATP~\footnote{ATP is the ``molecular unit of currency'' of intracellular energy transfer~\cite{KNO80} and is
used by many transporters that work against chemical gradients.} as an energy source~\cite{GUE09}, nevertheless trying to model
an ``energy consumption'' scenario is interesting to make some comparisons.
Since this is only a proof of concept there is no need to specify here in which form this energy is going to be provided, furthermore as long as we are only interested in comparing the initial rates of uptake we can avoid defining rules that regenerate energy in the cell. Therefore, rule R5 modelling the transporter role can be modified as follows:
\begin{equation*}\tag*{(R5')}
NH_4^+ \pipe \Loop{LjAMT2} \into (ENERGY\pipe X)\srewrites{k_t'}
H^+ \pipe \Loop{LjAMT2}\into (X \pipe NH_3\textrm{\_inside})
\end{equation*}
which consumes an element of energy within the cell. We also make
this reaction slower, since it is now catalysed by the concentration
of the $ENERGY$ element, actually, we set $k_{t'}=0.1*
10^{-10}$. Given the initial term $T=NH_3\times 138238 \pipe
NH_4^+ \times 138238 \pipe \Loop{LjAMT2}\into (ENERGY\times
100000)$ we obtain the simulation result in
Figure~\ref{fig:nh4_ljamt2_energy}. Note that the uptake work of the
transporter terminates when the $ENERGY$ inside the cell is
completely exhausted.

\subsection{$NH_4^+$ Diffusing from the Fungus}

We now model the diffusion of $NH_4^+$ from the fungus to the extracellular level (sections \textbf{(a1)}, and \textbf{(a2)} of Figure~\ref{fig:symbschem}). In section \textbf{(a1)} of the figure, the passage of $NH_4^+$ to the interfacial periarbuscular space happens by diffusion. We can model this phenomenon by adding a new compartment, representing the fungus, from which $NH_4^+$ flows towards the fungus-plant interface. This could be modelled through the rule:
\begin{equation*}\tag*{(R6)}
\Loop{FungMembr} \into (NH_4^+ \pipe X)\srewrites{k_f} NH_4^+ \pipe \Loop{FungMembr}\into X
\end{equation*}

\begin{figure}[t]
\begin{center}
\begin{minipage}{0.98\textwidth}
\begin{center}
\includegraphics[width=51mm]{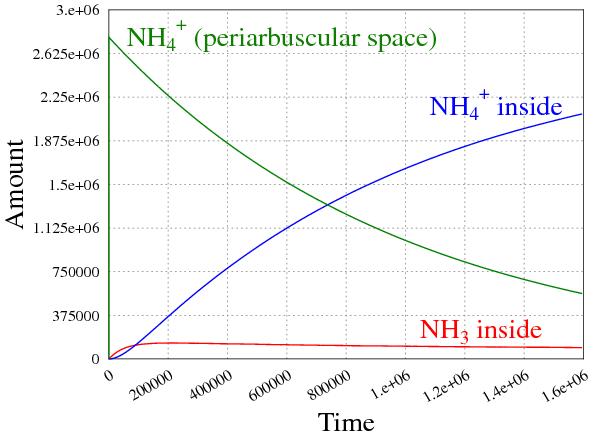}
\includegraphics[width=51mm]{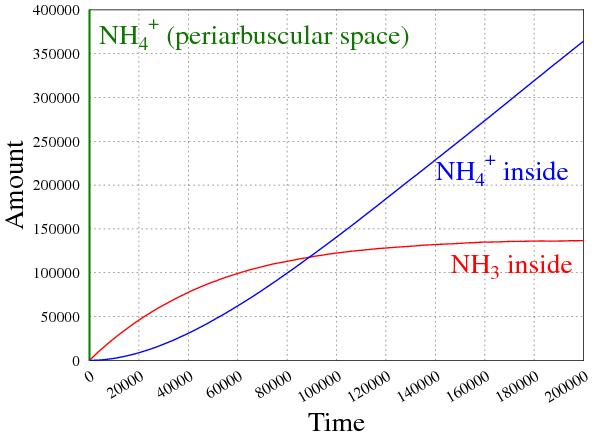}
\includegraphics[width=51mm]{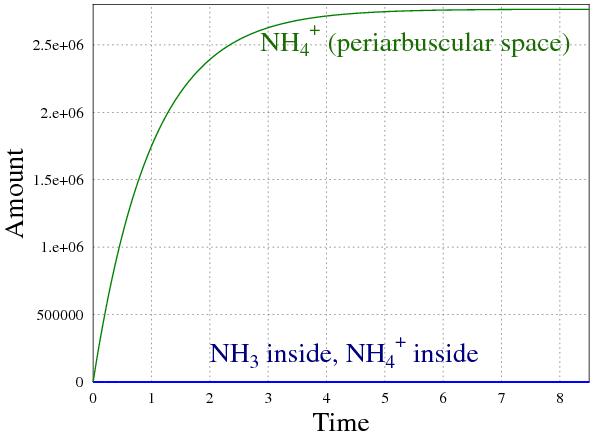}
\end{center}
\vspace{-0.5cm}
\caption{Diffusing $NH_4^+$ from the fungus, $k_f=1$.}
\label{fig:nh4_ljamt2_fungus}
\end{minipage}
\end{center}
\end{figure}

\begin{figure}[t]
\begin{center}
\begin{minipage}{0.98\textwidth}
\begin{center}
\includegraphics[width=51mm]{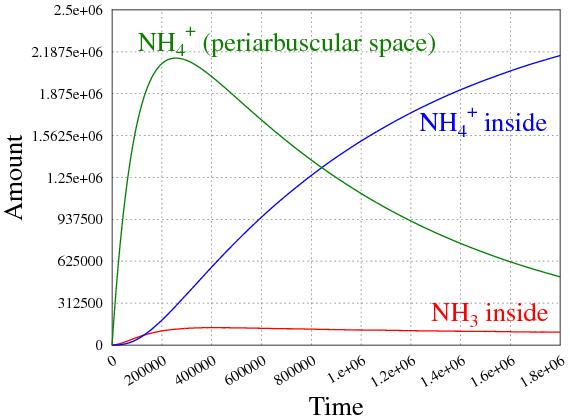}
\includegraphics[width=51mm]{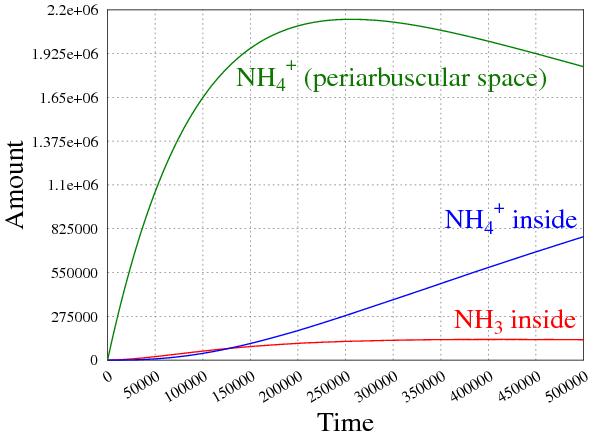}
\includegraphics[width=51mm]{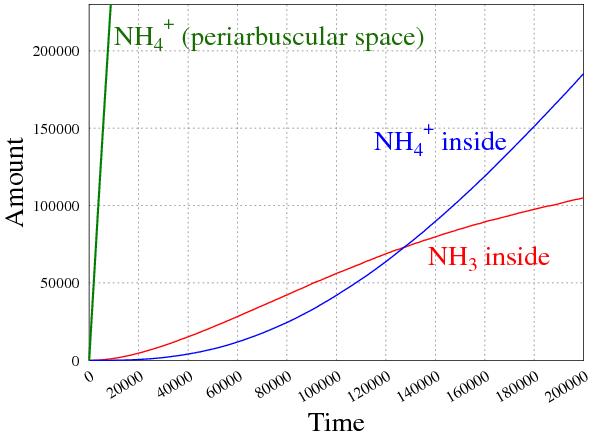}
\end{center}
\vspace{-0.5cm}
\caption{Diffusing $NH_4^+$ from the fungus, $k_f=0.01*10^{-3}$.}
\label{fig:nh4_ljamt2_fungus_slow}
\end{minipage}
\end{center}
\end{figure}

By varying the value of the rate $k_f$ one might model different
externalization speeds and thus test different hypotheses about
the underlying mechanism. In
Figure~\ref{fig:nh4_ljamt2_fungus} we give the simulation result,
with three different magnification levels, going from the whole simulation
on the left to the very first seconds on the right,
obtained from the initial term $T_f= \Loop{FungMembr} \into
(NH_4^+\times 2764677) \pipe \Loop{LjAMT2}\into \epsilon$ with
$k_f=1$. In the initial part, one can see how fast, in this case,
$NH_4^+$ diffuses into the periarbuscular space (in the figure,
$NH4$ represents the quantity of $NH_4^+$ in that part of the
system). In Figure~\ref{fig:nh4_ljamt2_fungus_slow} we give the
simulation result obtained from the same initial term $T_f$ with a
slower diffusion rate, namely $k_f=0.01*10^{-3}$ (note that the magnification levels in
the three panels are different with respect to those in Figure~\ref{fig:nh4_ljamt2_fungus}).

Finally, we would like to remark, without going into the simulation
details, how we can model in a rather natural way the portion
\textbf{(a2)} of Figure~\ref{fig:symbschem} in SCLS. Namely, we need
some rules to produce vesicle containing $NH_4^+$ molecules within
the fungal cell. Once the vesicle is formed, another rule drives its
exocytosis towards the interfacial space, and thus the diffusion of
the previously encapsulated $NH_4^+$ molecules. The needed rules are
given in the following:
\begin{equation*}\tag*{(R7)}
\Loop{FungMembr} \into ( NH_4^+ \pipe X)\srewrites{k_c} \Loop{FungMembr}\into (X \pipe \Loop{Vesicle}\into NH_4^+)
\end{equation*}
\begin{equation*}\tag*{(R8)}
\Loop{FungMembr} \into (NH_4^+ \pipe Y \pipe \Loop{Vesicle}\into X  )\srewrites{k_a} \Loop{FungMembr}\into (Y \pipe \Loop{Vesicle}\into (NH_4^+\pipe X)  )
\end{equation*}
\begin{equation*}\tag*{(R9)}
\Loop{FungMembr} \into (Y \pipe \Loop{Vesicle}\into X   )\srewrites{k_e} X \pipe \Loop{FungMembr}\into Y
\end{equation*}
Where rule R7 models the creation of a vesicle, rule R8 model the
encapsulation of an $NH_4^+$ molecule within the vesicle and rule
R9 models the exocytosis of the vesicle content.

%% file: SecDiscussion.tex
\section{Discussion}\label{SecDiscussion}
We dissected the route for the passage of $NH_3$ / $NH_4^+$ from the fungus to the plant in known and hypothetical mechanisms which were transformed in rules. Further, also the properties of the different compartments and their influence on the transported molecules were included, thus giving a first model for the simulation of the nutrients transfer.
With the model so far we can simulate the behaviour of the system when varying parameters as the different compartments pH, the initial substrate concentrations, the transport/diffusion speeds and the energy supply.

We can start comparing the two simulations with the plant cell
with the LjAMT2;2 transporter placed in different extracellular
situations: low $NH_4^+$ concentration
(Figure~\ref{fig:nh4_ljamt2_noNH4}) or high $NH_4^+$ concentration
(Figure~\ref{fig:nh4_ljamt2_noNH3}). As a natural 
consequence of the greater concentration, the ammonium
uptake is faster when the simulation starts with more $NH_4^+$, as
long as the LjAMT2;2 can readily import it. The real situation
should be similar to this simulation, assuming that the level of
extracellular $NH_4^+$ / $NH_3$ is stable, meaning an active
symbiosis. 

The simulation which represents an extracellular pH around 7 (Figure~\ref{fig:nh4_ljamt2_pH7}) shows a decreased internalisation
speed with respect to the simulation in Figure~\ref{fig:nh4_ljamt2_noNH3}, as could be inferred from the
concentrations of NH4\_inside and NH3\_inside in the plots on the right (focusing on the initial activity): this supports experimental data about
the pH-dependent activity of the transporter and suggests that the extracellular pH is fundamental
to achieve a sufficient ammonium uptake for the plants.
It's noticeable how the initial uptake rate in this case
is higher, despite the neutral pH, than the rate obtained considering an ``energy quantum''
used by the transporter (which has the same starting term), 
as could be seen in the right panels of Figure~\ref{fig:nh4_ljamt2_energy}
and Figure~\ref{fig:nh4_ljamt2_pH7}.
These results could enforce the biological hypothesis
%driven by protein structure analysis,
that, instead of ATP, a $NH_4^+$
concentration gradient (possibly created by the fungus) is used as
energy source by the LjAMT2;2 protein. 

The simulations which also consider the fungal counterpart are
interesting because they provide an initial investigation of this
rather
 poorly characterized side of the symbiosis and confirm that plants can efficiently gain
ammonium if $NH_4^+$ is released from the fungi.
This evidence supports the last biological hypothesis about how fungi supply nitrogen to plants~\cite{Gov05, Cha06},
and could lead to further models
which could suggest which is the needed rate for $NH_4^+$ transport from fungi to the interfacial apoplast;
thus driving biologists toward one (or some) of the nowadays considered hypotheses (active transport of $NH_4^+$, vesicle formation, etc.).

%% file: SecConc.tex
\section{Conclusions and Future Work}\label{SecConc}

This paper reports on the use of SCSL to simulate and understand
some biological behaviour which is still unclear to biologists,
and it also constitues a first attempt to predict biological
behaviour and give some directions to biologists for future
experiments.

SCLS has been particularly suitable to model the symbiosis (where
substances flow through different cells) thanks to its feature to
model compartments, and their membranes, in a simple and natural
way. Our simulations have confirmed some of the latest
experimental results about the LjAMT2;2 transporter~\cite{GUE09}
and also support some of the hypotheses about the energy source
for the transport. These are the first steps towards a complete
simulation of the symbiosis and open some interesting paths that
could be followed to better understand the nutrient exchange.

As demonstrated by heterologous complementation experiments in
yeast, mycorrhiza-specific plant transporters~\cite{Har02, GUE09}
show different uptake efficiencies under varying pH conditions.
Looking on this and the results from the simulations with
different pH conditions in the periarbuscular space new
experiments for a determination of the uptake kinetics
($K_m$-values) under a range of pH seems mandatory. Another
conclusion from the model is that, for accurate simulation, exact
in vivo concentration measurements have to be carried out even
though at the moment this is a difficult task due to technical
limitations.

As shown by various studies, many transporters on the plant side
show a strong transcriptionally regulation and the majority of
them are thought to be localized at the plant-fungus
interface~\cite{GUE09a}. Consequently a quantification of these
proteins in the membrane will be a prerequisite for an accurate
future model.

It is known that some of these transporters (e.g. PT4 phosphate
transporters) obtain energy from proton gradients established by
proton pumps (Figure~\ref{fig:symbschem}) whereas others as the
LjAMT2;2 and aquaporines~\cite{Gon06} use unknown energy supplies
or are simply facilitators of diffusion events along gradients.
Thus they conserve the membranes electrochemical potential for the
before mentioned proton dependent transport processes. To make the
story even more complex some of these transporters which are known
to be regulated in the AM symbiosis (e.g. aquaporines) show
overlaps in the selectivity of their substrates~\cite{Hol05,
JMZ04, Tye02, NT00}. Consequently integration of interrelated
transporters in a future model will be a necessary and challenging
task. It is quite probable that also transporters for other
macronutrients (potassium and sulfate) which might be localized in
the periarbuscular membrane influence the electrochemical
gradients for the known transport processes.

With the ongoing sequencing work~\cite{MGHL07} on the arbuscular
model fungus {\it Glomus intraradices} transporters on the fungal
side of the symbiotic compartment are likely to be identified and
characterized soon. Data from such future experiments could be
integrated in the model and help to answer the question whether
transporter mediated diffusion or vesicle based excretion events
lead to the release of ammonium into the periarbuscular
space~\cite{Cha06}.

Further questions about the plant nutrient uptake and competitive
fungal reimport process~\cite{Cha06, Bal07} might be answered.
Based on the transport properties of orthologous transporters from
different fungal and/or plant species, theories could be developed
which explain different mycorrhiza responsiveness of host plants;
meaning why certain plant-AM fungus combinations have a rather
disadvantageous than beneficial effect for the plant.

In future research on AM and membrane transport processes in
general values from measurements of concentrations or kinetics
can and have to be included in the model and will
show how the whole system is influenced by these values. Vice
versa simulations could be the base for new hypotheses and
experiments.
%Consequently simulated variations could help to
%assess which type of experiments might be the most promising ones
%to do. 
As a future extension of the modelling technique, we plan
to follow the direction taken in
\cite{DGT09,DBLP:journals/entcs/AmanDT09} using type systems to
guarantee that an SCLS term satisfies certain biological
properties (enforced by a set of typing rules), and also
investigate how type systems could enrich the study of
quantitative systems.